# Shear-induced unfolding and enzymatic cleavage of full-length VWF multimers

*Svenja Lippok[†], Matthias Radtke[‡], Tobias Obser[§], Lars Kleemeier[†], Reinhard Schneppenheim[§], Ulrich Budde[ǁ], Roland R. Netz[‡], and Joachim O. Rädler[†,*]*

[†] Faculty of Physics and Center for NanoScience, Ludwig Maximilian University, 80539 Munich, Germany

[‡] Department of Physics, Freie Universität Berlin, 14195 Berlin, Germany

[§] Department of Pediatric Hematology and Oncology, University Medical Center Hamburg-Eppendorf, 20246 Hamburg, Germany

[ǁ] Coagulation Lab, Medilys Laborgesellschaft mbH Hamburg, 22763 Hamburg, Germany

Running Title: Shear-induced unfolding and enzymatic cleavage of VWF

KEYWORDS: mechanosensitive proteins, polymer physics, enzyme kinetics, hemostasis, thrombotic microangiopathies

ABSTRACT

Proteolysis of the multimeric blood coagulation protein von Willebrand Factor (VWF) by ADAMTS13 is crucial for prevention of microvascular thrombosis. ADAMTS13 cleaves VWF within the mechanosensitive A2 domain, which is believed to open under shear flow. Here, we combine Fluorescence Correlation Spectroscopy (FCS) and a microfluidic shear cell to monitor real-time kinetics of full-length VWF proteolysis as a function of shear stress. For comparison, we also measure the Michaelis-Menten kinetics of ADAMTS13 cleavage of wild-type VWF in the absence of shear but partially denaturing conditions. Under shear, ADAMTS13 activity on full-length VWF arises without denaturing agent as evidenced by FCS and gel-based multimer analysis. In agreement with Brownian hydrodynamics simulations, we find a sigmoidal increase of the enzymatic rate as a function of shear at a threshold shear rate $\dot{\gamma}_{1/2}$ = 5522/s. The same flow-rate dependence of ADAMTS13 activity we also observe in blood plasma, which is relevant to predict hemostatic dysfunction.

INTRODUCTION

In the event of vascular injury, the multimeric plasma protein von Willebrand Factor (VWF) mediates platelet adhesion, aggregation, and crosslinking to maintain hemostasis [1,2]. VWF is either constitutively secreted into the circulation or is stored in Weibel-Palade bodies and released in the form of ultralarge multimers in response to physiological and pathophysiological stimuli [3]. A carefully controlled balance of VWF binding to platelets, and VWF cleavage by ADAMTS13 (a disintegrin and metalloprotease with thrombospondin type 1 motif, member 13) [4-6] is required for hemostasis. In this context, fluid shear stress plays an important role: It initially activates the binding of the VWF A1 domain to platelets, and regulates over time additionally VWF degradation by opening the A2 domain, exposing the cleavage site for ADAMTS13 [7,8]. Hence, force-controlled activation and cleavage of VWF is part of a feedback loop that balances hemostasis. Absence of ADAMTS13 causes the life-threatening disease thrombotic thrombocytopenic purpura (TTP), which is characterized by uncontrolled microvascular thrombosis [5,9]. ADAMTS13 may also control the size distribution of VWF in the circulation, since the effect of shear stress depends on the hydrodynamic drag and hence on the size of VWF multimers [10].

The mechanism of shear-induced unfolding of VWF has been explored in several studies in the recent past [11,12]. VWF is cleaved by breakage of the Tyr1605-Met1606 bond in its A2 domain, which is a deeply buried cleavage site [13,14]. The interaction of immobilized VWF or VWF fragments with ADAMTS13 fragments or variants has been characterized in buffer using ELISAs [7,15]. Recognition of recombinant VWF fragments by recombinant ADAMTS13 has been examined under both denaturing [16] and non-denaturing conditions [17]. In a seminal study by Zhang et al. using laser tweezers, force was applied directly to single A2 domains in buffer and the dependence of the catalytic rate on the enzyme concentration was measured [14]. Pulling at the ends of polymeric molecules results in a homogenous force profile along the polymer chain under equilibrium conditions. However, simulations of unfolding under shear flow reveal that the intramolecular forces in VWF are fluctuating and size dependent [18,19]. Therefore, cleavage probabilities under shear flow are expected to show a pronounced size dependence and more and less susceptible intramolecular regions may exist. In particular, ADAMTS13 is not expected to act on the full-length protein in the same fashion as on isolated fragments.

We have recently shown that Fluorescence Correlation Spectroscopy (FCS) allows one to measure the size distribution of VWF [20]. In addition, FCS can be used to monitor binding in complex fluids such as blood plasma [21] and provides information on macromolecular dynamics [22-24]. This is of particular interest for studies of VWF, since it is assumed that unidentified enhancers of ADAMTS13 activity are present in plasma, whose effects cannot be mimicked in experiments performed in buffer [25]. Moreover, since the sample size required is very small, FCS studies can easily be combined with microfluidic devices [26,27].

In this paper, we study the cleavage of VWF multimers by ADAMTS13 under shear flow in aqueous buffer as well as in blood plasma. Using FCS, we detect the breakdown of recombinant VWF-eGFP fusion protein (rVWF) by measuring the increase in molar VWF concentration. A specially designed shear cell, combined with a FCS setup, allows us to monitor the cleavage of full-length rVWF as a function of shear flow. With this approach, we analyzed VWF cleavage under denaturing conditions as a reference and the observed Michaelis-Menten constant was compared with that obtained for the disease-related mutant VWF-G1629E, which is abnormally sensitive to proteolysis [17]. We showed that cleavage of rVWF under shear flow in blood plasma can be followed over time. In normal plasma (NP), enhanced proteolytic activity was found, presumably due to hitherto unknown accessory factors. We then determined cleavage rates as a function of shear rate. In agreement with Brownian hydrodynamics simulations, the shear dependence is sigmoidal with a half-maximum shear rate $\dot{\gamma}_{1/2}$ = 5522/s. Theoretical modeling enabled us to calculate the probability of the A2 domain to be accessible to cleavage by ADAMTS13 from the distribution of tension along a coarse-grained polymer model of VWF.

RESULTS

**Cleavage of full-length rVWF follows Michaelis-Menten kinetics.** In this study, we adopted a novel approach to study VWF proteolysis, based on the use of recombinant VWF-eGFP fusion (rVWF) as substrate. Each VWF monomer is thus labeled with a single eGFP unit, and therefore cleavage of the rVWF-eGFP multimers results in an increase in the number of fluorescent molecules. We monitor this increase using the amplitude of the fluorescence correlation signal, G(0), which is inversely proportional to the average particle number N in the detection volume (see Figure 1a and Supporting Material S1). To test this concept, we first performed cleavage experiments in denaturing buffer (5 mM Tris, pH 8.0, with 10 mM BaCl$_2$, 1,5 M urea) that unfolds VWF in the absence of shear (Figure 1b-d). Figure 1b displays the decrease in G(0) with cleavage over time, which reflects an increase in the molar concentration of rVWF multimers (C$_{VWF}$). The derived rVWF concentration increased linearly in time, with the cleavage rate k$_{CR}$ = dC$_{VWF}$/dt depending on the ADAMTS13 concentration C$_A$ used (Figure 1c). For constant amounts of ADAMTS13, the cleavage rates increased hyperbolically as a function of the VWF concentration (Figure 1d). We fitted the curves with Michaelis-Menten kinetics

$$k_{CR} = k_{cat} C_A \cdot C_{VWF}^m / \left( K_M + C_{VWF}^m \right).$$  (1)

The substrate concentration $C^m_{VWF} = m\, C_{VWF}$ corresponds to the concentration of VWF monomers with the mean monomer number m = 5.6 (see Supporting Material S1). Imposing the constraint that the Michaelis-Menten constant $K_M$ does not depend on the concentration of ADAMTS13, we obtain the global Michaelis-Menten constant $K_M$ = (962 ± 480) nM and the catalytic rate constant $k_{cat}$ = (1.1 ± 0.4)×10$^{-3}$/s for the physiological ADAMTS13 concentration (1 µg/ml), designated here as 1 U/ml. A tenfold increase in protease concentration (10 U/ml) resulted in a reduced rate constant to $k_{cat}$ = (1.8 ± 0.6)×10$^{-4}$/s. The goodness of the fit for $k_{cat}$ is, however, limited, and this might explain the unexpected decrease in this parameter at the higher ADAMTS13 concentrations. Autoproteolysis of ADAMTS13 at high concentrations may also play a role.

**Cleavage of the mutant rVWF-G1629E.** Next, we asked whether ADAMTS13 cleaves disease-related VWF mutants with altered kinetics. We studied VWF-G1629E, which carries a mutation within the A2 domain near the site cleaved by ADAMTS13, as schematically depicted in Figure 2a. The G1629E mutation is believed to keep the domain permanently unfolded [17]. In agreement with this assumption, strong ADAMTS13 activity was detected even at zero shear. In order to compare the cleavage rates with those for wild-type rVWF, we plotted the mutant data together with the ADAMTS13 activity previously measured under denaturing conditions. Strikingly, the rates are comparable, although the enzyme concentration used in the mutant case is 200-fold lower (0.05 instead of 10 U/ml), (Figure 2b). However, it must be noted that the wild-type measurements were made under denaturing conditions, while the mutant was in buffer. For this reason we performed additional measurements comparing the proteolysis of the destabilized mutant rVWF-G1629E in buffer with and without urea. We found that the differences were not measurable within the experimental uncertainty (Supporting Material S2). Hence we assume that denaturing buffer has no strong impact on ADAMTS13 activity other than making the cleavage sites available in case of wild type VWF. This confirms that the mutant VWF is far more susceptible to proteolysis than the wild-type, leading to the loss of large multimers and ultimately to the significant bleeding symptoms seen in patients with the mutation G1629E.

**Shear-induced rVWF cleavage.** We studied the dependence of ADAMTS13-mediated rVWF cleavage under defined shear flow. To this end, a miniaturized shear cell was built that could be mounted on a microscope stage in combination with an FCS setup. We chose a Mooney-type design for the shear cell, consisting of a cylindrical container and a rotating inner cylinder with a conical tip, which generates constant shear rates throughout the sample volume of 150µl (see Figure 3a). This device allowed us to monitor enzyme activity under shear flow with shear rates up to 10.000/s. Since FCS measurements rely on diffusion, shearing and FCS data sampling had to be run in an alternating mode: The solution was sheared and VWF cleaved for 15 min, and this was followed by a 15-min FCS detection period in the absence of shear, during which the VWF concentration was determined (Figure 3a). A typical experiment consisted of 16 such cycles. Neglecting the very short time scale of ramping, we consider steady shear flow throughout the experiment. At zero shear rate, $\dot{\gamma}$ = 0/s, cleavage rates were very low, less than 2 pM/min. Thus, no appreciable change in VWF concentration is expected during the detection periods. However, higher shear rates did indeed result in increasing rVWF concentrations (Figure 3b). The full shear rate dependence was studied for cleavage of 10 nM rVWF at two different ADAMTS13 concentrations in buffer, with applied shear rates ranging from 0/s to 10.000/s (Figure 3c,d). The data show negligible enzymatic activity at low shear rates, with cleavage rates equivalent to that at zero shear, which is at the limit of resolution of the experiment. Significantly lower cleavage rates were detected for 1 U/ml ADAMTS13 compared to 10 U/ml. The increase in ADAMTS13 activity with shear can be described by different functional fits such as an exponential $dC_{VWF}/dt(\dot{\gamma}) = \alpha \cdot \exp(\dot{\gamma}/\dot{\gamma}_c)$, in which case a characteristic shear rate $\dot{\gamma}_c = (1090 \pm 44)$/s is obtained for both ADAMTS13 concentrations (Figure 3c,d dashed lines). A power law $dC_{VWF}/dt(\dot{\gamma}) = \alpha \cdot \dot{\gamma}^n$ equally well describes the data and results in the exponents $n_{U1} = 2.7$ and $n_{U10} = 3.6$ for the concentrations 1U/ml and 10U/ml, respectively. (see bold dotted lines in Figure 3c,d). At even higher shear rates, above 5000/s, an ill-defined regime appears, where FCS correlation curves indicate fluorescent objects with larger hydrodynamic radius rather than decreasing size. Since also the particle number concentration decreases in this regime, we assume that aggregation of the VWF multimers occurs. This process appears to be irreversible and the aggregates impede determination of reliable cleavage rates. Details of the course of aggregation, which might be due to shear-induced opening of intermolecular binding sites, are given in Supporting Material S3.

To corroborate the degradation of rVWF cleavage by ADAMTS13 under shear, we monitored the multimer distribution in the time course of a shear experiment using gel analysis (Figure 3e,f). Samples consisting of rVWF added to normal plasma (NP) were sheared at $\dot{\gamma} = 0$, 2700, 4000, 7500, and 10,000/s for 140 min. Gels were loaded with successive subsamples taken every 20 minutes. While no change in the multimer size distribution was observed up to a shear rate of $\dot{\gamma} = 2700$/s, large multimers disappeared with increasing shear rate above $\dot{\gamma} = 4000$/s (Figure 3e). Figure 3f exemplarily shows an enlarged view of the rapidly fading multimer size distribution as a function of shear time at the highest shear rate $\dot{\gamma} = 10,000$/s. Note that under the conditions in normal plasma we did not observe VWF aggregation at high shear rates, while in the measurements in buffer we observed the formation of aggregates for shear rates above 5000/s.

**Hydrodynamic Computer Simulations.** Our experiments show that rates of cleavage by ADAMTS13 in the regime of elevated shear rates very sensitively depend on the precise shear rate and that, in agreement with the general belief, mutations in the A2 domain can dramatically lower the shear rate at which cleavage becomes significant. In order to corroborate the concept of shear-induced opening of the A2 cleavage domain, we compared the measured shear-dependent cleavage rates with simulations of the tension profile in unfolded VWF. To this end, we modeled the VWF polymer as a chain of weakly aggregating beads connected by stiff springs and performed Brownian dynamics simulations including hydrodynamic interactions [18,28,29]. In our simulation model, spherical beads of radius a represent VWF's repeating units (dimers), which interact with a cohesive strength ε. Rescaled shear rates γτ were used, with the characteristic diffusion time of a single bead being given by τ = 6πηa³/kT. Thus, by appropriate rescaling the simulation results can be interpreted in terms of arbitrary values of radius and viscosity η. In order to obtain physical units the viscosity η = 1.2x10$^{-3}$ Pas and temperature T = 310K were used. The remaining parameters are the bead radius a and the cohesive strength ε, which strongly influence the shear rate at which shear-induced unfolding of the polymeric globule sets in. The cohesive parameter was set at ε = 2 kT, such that, in the absence of shear flow, the polymer is collapsed. The unfolding transition occurs at a rescaled shear rate $\dot{\gamma}\tau = 10$ [18], which translates to the experimental value of about 5000/s [11] when we set the bead radius to a = 73nm. Note that the unfolding transition is accompanied with maximal fluctuations of the polymer size due to shear-induced unfolding and refolding, as shown in the Supporting Material S4. We calculated tension profiles along the chain of beads under shear flow as a function of polymer size and shear rate (Figure 4). Figure 4a shows the profile of the average tension along the VWF polymer for a representative polymer size of 20 beads (tension profiles of 10mer and 30mer are depicted in Supporting Material S4). For the smallest shear rate, the force is maximal at the polymer termini. With increasing shear rate, the maximal forces move toward the chain center. Strong fluctuations are indicated by the high standard deviation σ, shown in the bottom of Figure 4a, and the corresponding broad tension distributions, shown for the middle bead i = 10 in Figure 4b. While fluctuations are larger than the average tensile forces (Figure 4a), the effect of increasing shear rate is more pronounced for the average force. Therefore, we analyze in the following the average tension profiles rather than fluctuations in order to study the shear-induced cleavage process. Typical snapshots of the 20mer unfolded by shear flow are illustrated in Figure 4c. In order to relate the tension profile to the cleavage activity, we calculate the equilibrium probability for a single subunit i to be open and accessible for cleavage by ADAMTS13 from the average tension $f_i(\dot{\gamma})$ between adjacent beads

$$P_i(\dot{\gamma}) = 1/(1 + \exp(-f_i(\dot{\gamma})/f_e)/K_0). \qquad (2)$$

This expression is derived considering the reversible transition of a cleavage site from the closed to the open state, where $K_0$ denotes the equilibrium constant in the absence of force and $f_e$ the effective force scale characterizing the force dependence of the probability $P_i$. As shown in the Supporting Material, the force scale $f_c$ of the closing transition dominates the effective force scale $f_e$ (S4, Data Analysis). Considering the entire multimer, we define the mean number of accessible cleavage sites, $n_{open}(N,\dot{\gamma}) = 2\sum_i^N P_i(\dot{\gamma})$ (two cleavage sites per dimer), and plot numerical results for different VWF dimer number N in Figure 4d for $f_e$ = 0.04 pN and $K_0$ = 0.0011. These parameters were obtained by mapping simulation results to the experimental data, as described below. In a phenomenological description, the shear dependence of $n_{open}$ is well approximated by a sigmoidal function (lines) where the saturation value is proportional to the VWF size 2N and the half-maximum shear rate decreases with increasing N. To take this size dependence into account, we weight $n_{open}$ according to the experimentally determined size distribution of VWF in blood plasma [20] and thereby obtain the size-average open probability

$$\bar{n}_{open}(\dot{\gamma}) = \sum_N b^{N-1} n_{open}(N,\dot{\gamma}) \Big/ \sum_N b^{N-1}, \qquad (3)$$

with base b = 0.64 (blue line, Figure 4d).

**Shear-induced cleavage in blood plasma.** We were particularly interested in characterizing the shear dependence of VWF proteolysis in blood plasma (Figure 5). Therefore, rVWF-eGFP was added to patient's blood plasma and studied in the FCS-shear cell. In this case, no ADAMTS13 is added, since blood plasma contains natural ADAMTS13. In this setting, both the recombinant VWF as well as the endogenous unlabeled VWF serve as substrates for ADAMTS13. We assume that there is no difference in the binding activity of ADAMTS13 for labeled and endogenous VWF because the binding site of the eGFP molecule is at the C-terminal and thus spatially separated from the cleavage site. Because VWF multimerization occurs only in vivo, no exchange is expected between labeled and endogenous VWF during the experiment. For data analysis, we estimate the concentration of unlabeled VWF to be 3 nM (see Supporting Material S1). Figure 5a shows the cleavage rate as a function of VWF concentration under denaturing conditions in 25% NP (0.25 U/ml ADAMTS13). For comparison, the plot also shows the data for 1 U/ml ADAMTS13 in buffer (taken from Figure 1d). Dotted lines indicate fits of Equation (1) with the same value for $K_M$ = 962 nM but different $k_{cat}$ = (6.4 ± 0.5)×10$^{-3}$/s (NP) and $k_{cat}$ = (1.1 ± 0.4)×10$^{-3}$/s (buffer). The patient's plasma used in Figure 5b contained anti-ADAMTS13 autoantibodies (AB plasma) and cleavage rates are considerably reduced ((3.13 ± 1.5)pM/min instead of (18.24 ± 2.0)pM/min). Figure 5c shows the cleavage rate as a function of shear. Importantly, in agreement with the gel analysis shown in Figure 3e,f, we did not find any evidence for VWF aggregation in experiments with blood plasma. We were therefore able to measure VWF cleavage in blood plasma up to $\dot{\gamma}$ = 10.000/s. In order to exclude the possibility that VWF concentration might increase under high shear over time due to solvent evaporation, we carried out a control experiment on pure eGFP protein, which shows only a small concentration increase of 6% in buffer subjected to a shear rate of $\dot{\gamma}$ = 8400/s over 2.5h. Under shear, rVWF in 50% NP exhibits a sigmoidal increase in cleavage rates with increasing shear rates. For the 50% AB plasma with reduced ADAMTS13 activity, we find the same sigmoidal dependence on shear, but overall reduced cleavage rates. The dotted lines in Figure 5c are best fits with a phenomenological equation for the shear-dependent cleavage rate

$$k_{CR}(\dot{\gamma}) = k_{max} \big/ \big(1 + \exp(-(\dot{\gamma} - \dot{\gamma}_{1/2}) / \Delta\dot{\gamma})\big),  \qquad (4)$$

where $\dot{\gamma}_{1/2}$ denotes the half-maximum shear rate, at which half of the monomers are accessible to cleavage, and the parameter $\Delta\dot{\gamma}$ determines the width of the transition. The values for $k_{max}$ denote the maximal enzymatic rate in the case of fully accessible cleavage sites. These will depend on the ADAMTS13 concentration $C_A$ as described by the Michaelis-Menten kinetics (Equation (1)). We obtain as a best fit the half maximum shear rate $\dot{\gamma}_{1/2}$ = (5522 ± 642)/s and width $\Delta\dot{\gamma}$ = (1271 ± 371)/s and enzyme activity dependent prefactors $k_{max}$ = (3.5 ± 0.4)×10$^{-3}$ nM/s and $k_{max}$ = (1.4 ± 0.3)×10$^{-3}$ nM/s for NP and AB plasma, respectively (Figure 5c, dotted lines). Notice that the half maximum shear rate is close to the critical shear rate of VWF unfolding (5000/s [11]) indicating the correlation between high cleavage activity and large polymer size fluctuations. The fact that the shear dependence of cleavage in both NP and AB plasma has the same shape is a strong indication that VWF A2 domain opening is the rate limiting mechanism, while the conditions in the blood plasma, primarily the concentration of available ADAMTS13, sets the maximal enzymatic rate.

As seen in Figure 5c, the experimentally determined dependence of ADAMTS13 activity on shear is in good agreement with the hydrodynamic model. The central assumption that we make is that the cleavage rate $k_{CR}$ depends on the shear-dependent mean number of accessible cleavage sites $\bar{n}_{open}$ via the effective substrate concentration

$$C^m_{VWF} = \bar{n}_{open} C_{VWF}  \qquad (5)$$

that enters the Michaelis-Menten Equation (1), where $\bar{n}_{open}$ is given by Equation (3). This is in contrast to measurements under denaturing conditions where the mean number of monomers m = 5.6 was assumed. For the theoretical prediction, the same value for $K_M$ = 962 nM is used. Given fixed concentrations $C_A$ = 1.3 nM and $C_{VWF}$ = 13 nM, the simulation results for the cleavage rate (open blue symbols in Figure 5c) fit the experimental data for NP plasma (filled blue symbols) for adjusted fit parameters $f_e$ = 0.04 pN, $K_0$ = 0.0011, and the catalytic rate constant for a single cleavage site $k_{cat}$ = 0.04/s. Notice that the length scale $x_e kT/f_e \approx$ 100 nm is comparable to the double contour length of the A2 domain [14]. Using the same fit parameters for AB plasma (black symbols), we obtain as a best fit an effective ADAMTS13 concentration $C_A$ = 0.55 nM. Note that $k_{cat}$ = 0.04/s obtained in shear-dependent measurements (Figure 5c) is one order of magnitude larger than $k_{cat}$ = 0.0064/s obtained under denaturing conditions (Figure 5a).

DISCUSSION

Using FCS and recombinant rVWF-eGFP, we have measured the ADAMTS13 activity in real-time. In denaturing buffer, we observed Michaelis-Menten kinetics with $K_M$ = 962 nM. This value is in good agreement with those obtained in recent experiments performed with VWF fragments: Zanardelli et al. measured $K_M$ = 1.61 $\mu$M ($k_{cat}$ = 0.14/s) by high pressure liquid chromatography for cleavage of a 16.9 kDa fragment of the VWF A2 domain [30]. In experiments using gel analysis and ELISA, Gao et al. obtained $K_M$ = 1.7 $\mu$M ($k_{cat}$ = 1.3/s) for cleavage of an A2 domain fragment consisting of amino acid residues Asp1596-Arg1668 [7]. Using a VWF fragment suited for fluorescence energy transfer studies, Di Stasio et al. measured $K_M$ = 4.6 $\mu$M ($k_{cat}$ = 0.03/s) at pH 6.0 [31]. The differences to the catalytic rate constants that we observed, $k_{cat}$ = 0.0011/s, and 0.00018/s, depending on the ADAMTS13 concentrations, presumably are related to the fact that the earlier studies investigate cleavage independently of substrate unfolding in presumably less physiological situations. Thus their values seem to reflect higher proteolytic activities than found in our work. However, our data, for the first time, yield rates of full-length VWF cleavage.

The unfolding of VWF under shear is known as a globular-stretch transition and it has long been speculated that its susceptibility to proteolysis is modulated by shear-induced extension. For the cleavage experiments under shear, we used a custom-built shear cell. Multimer analysis proved that the size distribution of the sheared VWF sample shows the same characteristics than VWF samples cleaved in a static assay under denaturing conditions [32]. Using FCS, we quantified the shear dependence of the VWF breakdown exhibiting a sigmoidal dependence on shear with half maximum shear rate $\dot{\gamma}_{1/2}$ = 5522/s and width $\Delta\dot{\gamma}$ = 1271/s. These findings are in line with earlier observation of abrupt onset of VWF unfolding as a function of shear rate using fluorescence microscopy and small-angle neutron scattering [11,12]. Our Brownian hydrodynamics simulations show that hydrodynamic shear flow is capable to unfold VWF multimers and suggest that the internal tension opens the A2 domains in a dynamic equilibrium. We calculated the probability of a single subunit to be open as a function of the average tensile force $f_i$ acting between two neighboring monomers. Comparing the simulation to experimental data yields an effective force scale $f_e$ = 0.04 pN, at which the A2 domain opening probability begins to increase. This value should be compared to the characteristic force scale of A2 unfolding, $f_\beta$ = 1.1pN, previously measured by force-induced unfolding of the isolated A2 domain using optical tweezers [14]. In that study, unfolding forces were measured as a function of force loading rates. Clearly, the force at which A2 is fully unfolding as probed in force spectroscopy experiments is not the same force as required to render A2 domains accessible to ADAMTS13 cleavage under shear [8,14]. In fact, $f_e$ is found to be substantially smaller than $f_\beta$. It is interesting to note that in our two-state equilibrium model the force dependence of the opening and closing rates are different, and that the force scale of the closing rate dominates $f_e$, which corroborates the difference between $f_e$ and $f_\beta$ (see Supporting Material S4). Also Zhang et al. [14] determined the rate of A2 refolding. We compare the rates in the absence of force and thus the free energy difference between the two states. As a result, we obtain an estimate for the free energy difference $\Delta F = -\ln(K_0) kT = 6.8$ kT, which is in good agreement to the single barrier A2 unfolding and refolding kinetic model (6.6 kT) [14]. We consider this agreement surprising given the slightly different model assumptions that go into the analysis of the experimental data. In terms of enzyme activity, Zhang et al. determined a catalytic rate constant of 0.14/s for single, accessible A2 domains and varying enzyme concentration [14]. Since we measured shear-induced cleavage of full-length VWF in blood plasma instead of unfolded fragments in buffer, the smaller rate constant that we obtained, $k_{cat} = 0.04$/s, might not be surprising.

Our study of the VWF proteolysis induced by shear flow represents closely the clinical relevant situation. Our model does not require a mechanical force-induced unfolding of the A2 domain but rather considers the cleavage in the context of a thermodynamic reversible two-state opening model, in which a full-length VWF molecule is subject to shear flow and hence includes size dependent effects of VWF cleavage. Since both VWF size distribution as well as local mutations in VWF's A2 domain are known to affect the ADAMTS13 susceptibility, it is desirable to extend the shear-induced-opening model even further in order to bring atomistic models together with coarse-grained hydrodynamic model used in this work. In the future such approaches might be capable to relate mutations directly to hemostatic dysfunction.

A second aspect of the work is the measurement of ADAMTS13 activity in its natural environment, namely blood plasma, which is of clinical importance. We determined the kinetics of rVWF-eGFP cleavage in 25% normal blood plasma and found that the rate of proteolysis under shear is increased relative to experiments performed in buffer. This finding corroborates a previous hypothesis, which suggested that certain components found in blood, such as coagulation factor VIII, increase VWF's susceptibility to cleavage [33]. Although, we did not study the effect of particular blood plasma components on our studies, it is known that the enhancement of proteolysis is attributed, inter alia, to the conformational change induced in VWF by the binding of factor VIII [34]. Furthermore, calcium is known to have a strong down-regulating effect on VWF cleavage due to calcium-mediated stabilization of the A2 [35]. Other cofactors and their effects on VWF have not yet been studied, but might provide novel therapeutic targets. As a consequence, FCS-based measurements of rVWF-eGFP cleavage may have diagnostic potential. Determination of levels of ADAMTS13 activity in plasma is, for example, required for differential diagnostics in microangiopathies. FCS combined with recombinant VWF-eGFP added to patient samples permits detection of extremely low levels of ADAMTS13 activity. Even detection limits well below 5% can be achieved, if the extremely proteolysis-sensitive rVWF-G1629E is employed as a substrate. Furthermore, FCS can help to quantify the effect of VWF mutations and assess the impact of potential drugs. In this context, the ability to measure specific cleavage activity on VWF in native blood samples constitutes a valuable extension of available methods. In summary, quantitative assessment of shear-induced enhancement of ADAMTS13 activity, both in vitro and in patient's blood samples, together with the fact that the data follow theoretical modeling, provide the means to shed light on the new territory of shear-dependent phenotypes in clinically relevant VWF mutations.

MATERIALS AND METHODS

**Protein expression:** The recombinant fusion protein rVWF-eGFP was expressed and purified as described in detail before [20]. For the mutant VWF-G1629E, the plasmid PIRESneo2-VWF-G1629E-eGFP was used, which was produced by in vitro mutagenesis of wild-type pIRESneo2-VWF-eGFP by replacing wild-type codon 1629 for glycine with the naturally occurring mutant codon specifying glutamic acid (p.Gly1629Glu). Wild-type rhuADAMTS13 used was produced as described in [17].

**Plasma samples** were taken from normal volunteer donors and collected in S-monovettes, coagulation sodium citrate (Sarstedt, Germany), incubated at room temperature (30 min), and centrifuged (10 min, 2300 rpm). The supernatant was aliquoted and stored at −80°C until use. Informed consent was obtained from all subjects. VWF concentration was measured to be 97%, i.e. in the normal range by an ELISA measuring collagen-binding capacity (MEDILYS Laborgesellschaft mbH, Hamburg, Germany). ADAMTS13 activity was obtained by an activity-ELISA to be 113% (Technoclone, Vienna, Austria).

**Fluorescent Correlation Spectroscopy (FCS)** measurements were performed on an Axiovert 200 microscope with a ConfoCor 2 unit (Carl Zeiss, Jena, Germany), equipped with a 40x (NA = 1.2) water-immersion Apochromat objective (Carl Zeiss). A 488 nm argon laser was used for illumination. For calibration, eGFP was measured for 20x60 s in the corresponding buffer. Experiments without shear flow were carried out in eight-well LabTek I chamber slides (Nunc, Rochester, NY) on rVWF for 20x15 min and on rVWF-G1629E for 20x9 min, (owing to its higher sensitivity to proteolysis). Correlation analysis was performed using the ConfoCor 2 software. The FCS data analysis is described in detail in Supplementary Information S1.

**Shear cell:** A custom-made shear cell was used to apply shear forces consisting of a combined cone-and-plate and concentric cylindrical design, called Mooney system [36]. A coverslip (Carl Roth, Karlsruhe, Germany) was fixed with nail varnish to a stainless steel sample holder containing wells of radii $r_o$ = 6.5 mm, serving as outer cylinders. The inner rotating cylinder (radius $r_i$ = 6.1 mm) was connected to a brushless motor with integrated controller (Faulhaber GmbH, Schönaich, Germany). Before use, both cylinders were first incubated for 30 min with UHT milk to avoid sticking of the sample to the surface, and then washed with buffer for at least ten times. For shear-flow measurements, periods of applied shear stress (15 min) without FCS recording alternated with measuring periods (5x3 min) without shear flow.

**Multimer analysis** was performed by sodium dodecyl sulfate agarose gel electrophoresis in combination with immunoblotting and luminescent visualization [37]. The luminescent blot was stored on electronic media using photo imaging (FluorChem8000; Alpha Innotech, San Leandro,

CA) [38]. The sample consisted of 3 nM rVWF-eGFP in 5 mM Tris-HCl (pH 8.0) and 50% normal blood plasma.

**Cleavage protocol:** Measurements were conducted in either buffer or blood plasma. The cleavage buffer contained 5 mM Tris-HCl (pH 8.0) with 10 mM $BaCl_2$ for ADAMTS13 activation [31,39]. For measurements without shear flow, 1.5 M urea was added to partially denature VWF based on the commonly used protocol [30,39,40]. For measurements in blood plasma, the plasma was diluted to the desired concentration with 5 mM Tris-HCl (pH 8.0). For all measurements, the temperature was set to 37°C (without shear: heating stage, ibidi GmbH, Martinsried, Germany; with shear: water bath, Julabo GmbH, Seelbach, Germany).

**Simulation methods:** Brownian dynamics simulations were performed using a discretized Langevin equation, where hydrodynamic interactions are taken into account via the Rotne-Prager tensor [41] (for details, see Supporting Material S4). The VWF was modeled as a homopolymer consisting of N beads, which interact via Lennard-Jones potentials of depth $\varepsilon = 2$ kT, and are connected in a linear chain by harmonic bonds with a rescaled spring constant $k = 200$ kT/$a^2$. Tensile forces acting along each bond $f_i = k(r_{i,i+1} - 2a)$ are recorded during the course of simulation. To compare the dimensionless quantities used in the simulation to physical values, we use the dimer radius $a = 73$ nm, the viscosity $\eta = 1.2 \times 10^{-3}$ Pas, and the temperature $T = 310$ K.

FIGURES

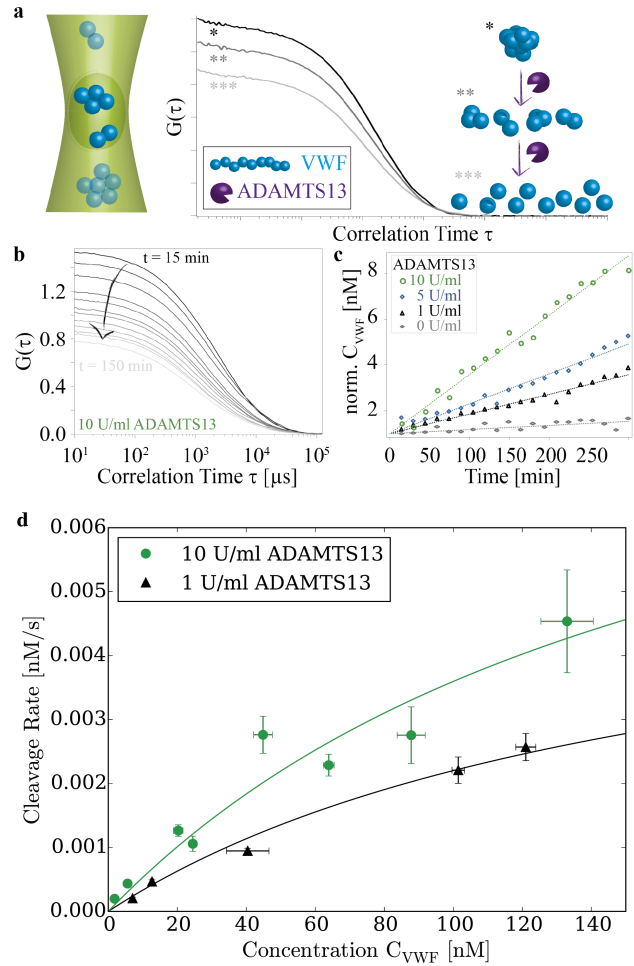

**Figure 1.** FCS analysis of the kinetics of full-length rVWF proteolysis in buffer. (a) The correlation curve G(τ) at correlation time 0 is proportional to the reciprocal of the number of fluorescent molecules in the detection volume N = 1/G(0), i.e. the molar concentration of VWF. For multimeric rVWF-eGFP with one eGFP molecule attached to each VWF monomer, cleavage by ADAMTS13 results in an increase in the number of fluorescent molecules. This increase in N concomitantly reduces the value of G(0) and can be detected in real-time. (b) Representative autocorrelation curves showing that proteolysis of VWF is reflected in a progressive decrease in amplitude G(τ) with time. (c) Molar concentration of rVWF molecules increases linearly over time and the cleavage rates depend on ADAMTS13 concentration. (d) The hyperbolic dependence of cleavage rate on VWF concentration was fitted with Equation (1) (lines). Constraining a Michaelis-Menten constant independent of the enzyme concentration, we obtained $K_M$ = 962 nM, and $k_{cat}$ = 0.0011/s and $k_{cat}$ = 0.00018/s for $C_A$ = 1 U/ml and $C_A$ = 10 U/ml, respectively.

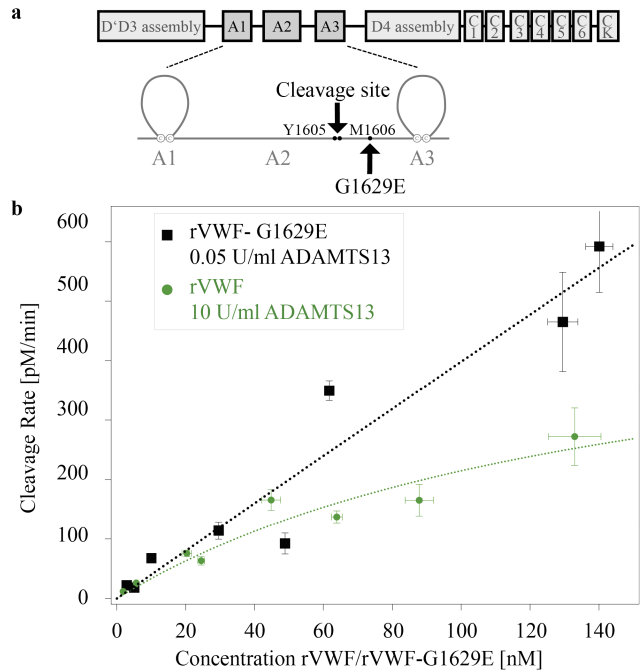

**Figure 2.** Proteolysis of the disease-related mutant rVWF-G1629E. (a) Schematic drawing of the multidomain structure of a VWF monomer. The G1629E mutation is located close to the ADAMTS13 cleavage site (Y1605-M1606) and results in constitutive opening of the A2 domain. (b) Cleavage kinetics of rVWF-G1629E in the absence of shear. In buffer without urea, the mutant exhibits significantly increased susceptibility to proteolysis as compared to rVWF in denaturing buffer. Note that cleavage rates in both cases are comparable, even though a 200-fold lower ADAMTS13 concentration (0.05 U/ml) was used in case of the mutant. Best Michaelis-Menten kinetic fits are shown as dotted lines.

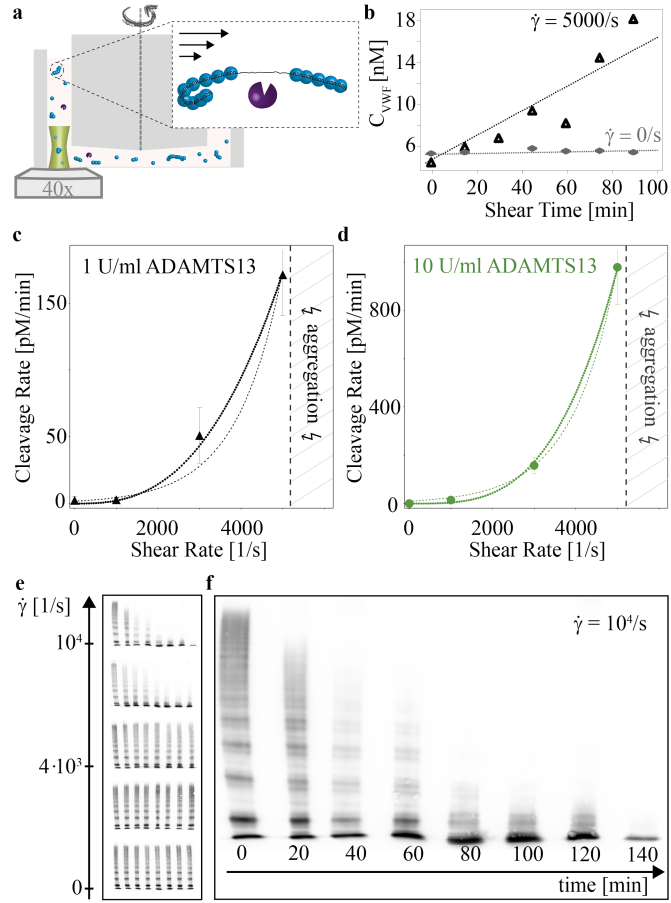

**Figure 3.** Shear-induced cleavage of rVWF. (a) Cartoon illustrating VWF proteolysis under shear flow. The shear cell generates constant shear rates throughout the whole sample volume. Above a certain threshold, VWF unfolds, the A2 domains open and cleavage sites for ADAMTS13 are exposed. (b) Molar VWF concentrations at the end of each shear period are plotted as a function of the total duration of shear as shown here for $\dot{\gamma}$ = 0/s and 5000/s (1 U/ml ADAMTS13). (c) rVWF cleavage rates (1 U/ml ADAMTS13, 10 nM rVWF) as a function of applied shear rates. (d) Use of a tenfold higher protease concentration (10 U/ml) significantly increases cleavage rates. Exponential (dashed lines) and power law (bold dotted lines) fits are plotted. For shear rates $\dot{\gamma}$ > 5000/s, VWF aggregation becomes dominant over VWF cleavage. (e) Multimer analysis of rVWF proteolysis in 25% NP under shear. Miniaturized gel data visualize the effect of increasing cleavage with increasing shear rate (bottom to top: $\dot{\gamma}$ = 0, 2700, 4000, 7500, and 10,000/s). (f) Enlargement of the first plot depicts the time dependence of VWF breakdown at $\dot{\gamma}$ = 10,000/s. Note that performing the cleavage experiment in blood plasma prevents the formation of VWF aggregates.

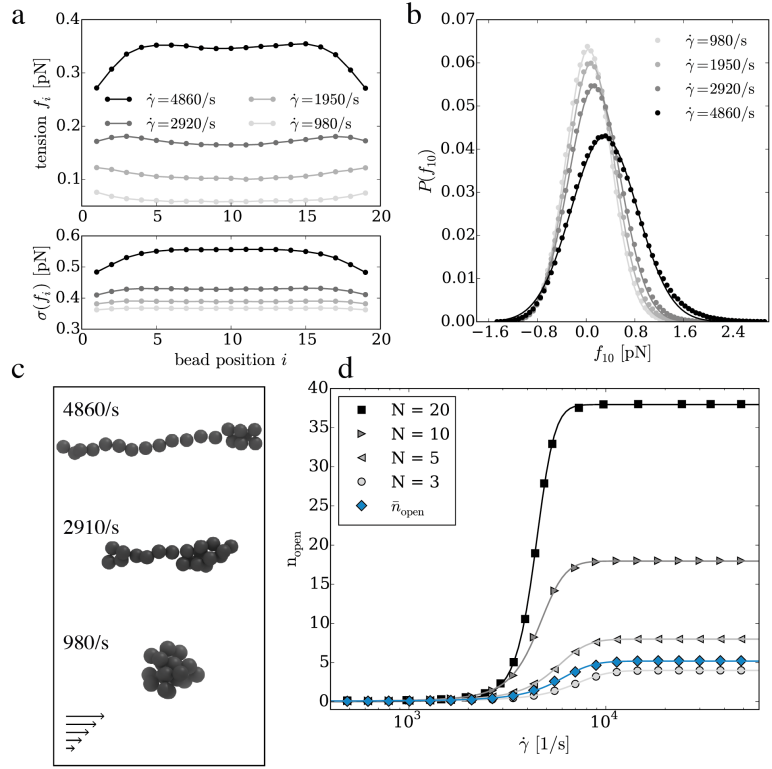

**Figure 4.** Results of hydrodynamic simulations for the polymer response to shear flow. (a) The average tension $f_i$ is plotted as a function of bead position $i$ along a polymer of size $N = 20$. The tension exhibits a complex profile that increases with increasing shear rate $\dot{\gamma}$. The standard deviation $\sigma(f_i)$, plotted at the bottom, indicates a broad distribution $P(f_i)$ of the tensile force, which are shown in (b) for the middle bead $i = 10$. (c) Snapshots of typical polymer configurations obtained at different shear rates as indicated. (d) The mean number of open cleavage sites $n_{open}$ is plotted as a function of shear rate for different multimer sizes $N$. It exhibits a sigmoidal increase with increasing shear rates; sigmoidal fits according to Equation (4) are indicated by solid lines. $\bar{n}_{open}$ denotes the weighted average according to the exponentially distributed VWF multimer sizes (blue line).

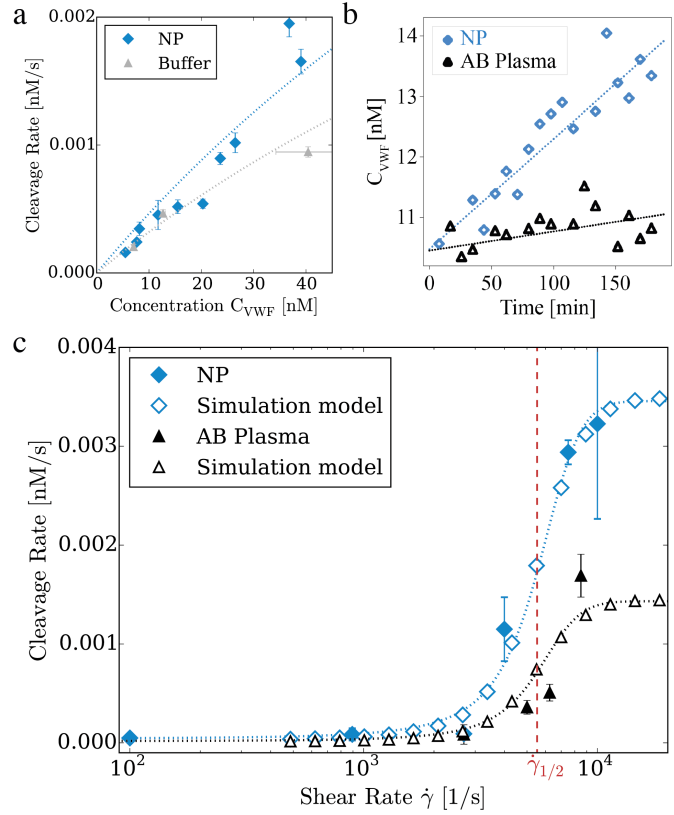

**Figure 5.** VWF proteolysis in blood plasma. (a) Kinetic analysis of rVWF cleavage in 25% NP (0.25 U/ml ADAMTS13) under denaturing conditions. Fitting Equation (1) (blue line) for fixed $K_M$ = 962 nM yields $k_{cat}$ = 0.0064/s. For comparison, the (lower) enzyme activity in buffer (1 U/ml ADAMTS13) is shown (grey). (b) Protease activity in normal versus patient's AB plasma. The patient's plasma exhibited a six-fold reduced protease activity under denaturing conditions. (c) Shear-flow induced cleavage in blood plasma. For both NP and AB plasma, cleavage rate increases following a sigmoidal function, Equation (4), with increasing shear ($\dot{\gamma}_{1/2}$ = 5522/s, $\Delta\dot{\gamma}$ = 1271/s, $k_{max}$ = 0.0035 nM/s and $k_{max}$ = 0.0014 nM/s for NP and AB respectively, fit: dotted lines) in good agreement with the Brownian hydrodynamic simulation assuming that the cleavage rate is described by Equation (1) with concentration of accessible VWF monomers given by Equations (5),(3), and (2). (dotted lines and open markers).

ASSOCIATED CONTENT

**Supporting Material**. S1 Analysis of VWF enzymatic degradation using FCS, S2 The impact of denaturing conditions on ADAMTS13 activity and eGFP, S3 VWF aggregation at high shear rates exceeds ADAMTS13 cleavage, S4 Brownian dynamic simulation.

AUTHOR INFORMATION

**Corresponding Author**

\* raedler@lmu.de

**Author Contributions**

J.O.R., R.R.N., and R.S. designed the study. S.L. and L.K. performed the FCS experiments. M.R. performed the MD simulations. T.O. engineered the recombinant proteins. S.L. and U.B. performed the multimer analysis. S.L., R.R.N., and J.O.R. analyzed and interpreted the data. S.L., M.R., and J.O.R. wrote the manuscript. R.S., R.R.N., and U.B. critically revised the manuscript. All authors have given approval to the final version of the manuscript.

ACKNOWLEDGMENT

We thank Thomas Ligon for reading the manuscript. This work was supported with seed funds from the Center for NanoScience, and by DFG research unit FOR 1543 (www.shenc.de), which is funded by the Deutsche Forschungsgemeinschaft. We gratefully acknowledge all SHENC members, especially Frauke Gräter for fruitful dis-cussions. S.L. thanks the Elite Network of Bavaria for its support.

SUPPORTING CITATIONS

References (42-47) appear in the Supporting Material.